\documentclass[english, 12pt]{amsart}
\usepackage[T1]{fontenc}
\usepackage[latin9]{inputenc}
\usepackage{amsthm}
\usepackage[numbers]{natbib}

\makeatletter
\numberwithin{equation}{section}
\numberwithin{figure}{section}
\theoremstyle{plain}
\newtheorem{thm}{Theorem}
  \theoremstyle{definition}
  \newtheorem{defn}[thm]{Definition}
  \theoremstyle{plain}
  \newtheorem{cor}[thm]{Corollary}
  \theoremstyle{plain}
  \newtheorem{prop}[thm]{Proposition}
  \theoremstyle{plain}
  \newtheorem{lem}[thm]{Lemma}
  \theoremstyle{remark}
  \newtheorem*{rem*}{Remark}

\usepackage{amssymb}

\makeatother

\usepackage{babel}

\begin{document}

\title{Three-parameter complex Hadamard matrices of order 6.}

\author{Bengt R. Karlsson}

\address{Uppsala University, Dept of Physics and Astronomy, Box 516, SE-751
20, Uppsala, Sweden}

\email{bengt.karlsson@physics.uu.se}
\begin{abstract}
A three-parameter family of complex Hadamard matrices of order 6 is
presented. It significantly extends the set of closed form complex
Hadamard matrices of this order, and in particular contains all previously
described one- and two-parameter families as subfamilies. 
\end{abstract}
\maketitle

\section{Introduction}

Complex Hadamard matrices have turned out hard to classify, with current
classifications being incomplete for order 6 and higher. For order
6, there is evidence for a four-parameter family \citep{Bengtsson,Skinner},
but up till now only zero-, one- and two-parameter subfamilies have
been obtained on closed form, as reviewed in \citep{Tadej guide,Tadej webguide}.
Recent progress includes the construction of three two-parameter,
nonaffine families \citep{Szollosi,Karlsson_JMP} that contain the
one-parameter families as subfamilies, and has resulted in an overall
picture of five, partially overlapping, two-parameter families of
complex Hadamard matrices of this order. 

A further step towards a more comprehensive classification was taken
in \citep{Karlsson II}, where it was shown that any complex Hadamard
matrix of order 6 is equivalent to (or equals) a Hadamard matrix for
which either all (the \hbox{$H_2$-reducible} case) or none of its
nine $2\times2$ submatrices are Hadamard. In the present paper, a
complete characterization of the $H_{2}$-reducible Hadamard matrices
is given. The result is a three-parameter family which has all the
previously known (one- and) two-parameter families as subfamilies.

\section{Preliminaries}

An $N\times N$ matrix $H$ with complex elements $h_{ij}$ is Hadamard
if all elements have modulus one, $|h_{ij}|=1$, and if\begin{equation}
HH^{\dagger}=H^{\dagger}H=NE\label{eq:unitarity}\end{equation}
(the unitarity constraint), where $E$ is the unit matrix in $N$
dimensions. Two Hadamard matrices are termed equivalent, $H_{1}\sim H_{2}$,
if they can be related through\begin{equation}
H_{2}=D_{2}P_{2}H_{1}P_{1}D_{1}\end{equation}
where $D_{1}$ and $D_{2}$ are diagonal unitary matrices, and $P_{1}$
and $P_{2}$ are permutation matrices. A set of equivalent Hadamard
matrices can be represented by a dephased matrix, with ones in the
first row and the first column. 

The present paper will be concerned with Hadamard matrices which are
reducible in the following sense.
\begin{defn}
A complex Hadamard matrix of order 6 is $H_{2}$-reducible if it is
equivalent to a Hadamard matrix for which all the nine $2\times2$
submatrices are Hadamard. 
\end{defn}
$H_{2}$-reducible Hadamard matrices are more prevalent than might
be thought. The quite general nature of these matrices is illustrated
by the following theorem that was proven in \citep{Karlsson II}.
\begin{thm}
\label{thm:Karlsson II}Let $H$ be a complex Hadamard matrix of order
6, with elements $h_{ij}$, $i,j=$1,...,6. If there exists an order
2 submatrix $\left(\begin{array}{cc}
h_{ij} & h_{ik}\\
h_{lj} & h_{lk}\end{array}\right)$ that is Hadamard, then $H$ is $H_{2}$-reducible.
\end{thm}
Since the submatrix referred to in Theorem \ref{thm:Karlsson II}
has the (unique) dephased form\begin{equation}
F_{2}=\left(\begin{array}{cc}
1 & 1\\
1 & -1\end{array}\right),\label{eq:F2}\end{equation}
$H_{2}$-reducible Hadamard matrices are easily identified:
\begin{cor}
Let H be a complex Hadamard matrix of order 6. H is $H_{2}$-reducible
if, and only if, its dephased form has at least one element equal
to -1. 
\end{cor}
It follows from the corollary that all the currently known one- and
two-parameter families in six dimensions ($F_{6}^{(2)},\,(F_{6}^{(2)})^{T}$,$D_{6}^{(1)}$
\citep{Dita}, $B_{6}^{(1)}$ \citep{Beau_Nic}, $M_{6}^{(1)}$ \citep{Mat Szo},
$X_{6}^{(2)},\,(X_{6}^{(2)})^{T}$ \citep{Szollosi} and $K_{6}^{(2)}$
\citep{Karlsson_JMP}, in the notation of \citep{Tadej guide,Tadej webguide})
are families of $H_{2}$-reducible Hadamard matrices. On the other
hand, the single, isolated matrix $S_{6}^{(0)}$ is not $H_{2}$-reducible.

A general, $H_{2}$-reducible Hadamard is equivalent to a Hadamard
matrix on the dephased form (see \citep{Karlsson II})\begin{equation}
H=\left(\begin{array}{ccc}
F_{2} & Z_{1} & Z_{2}\\
Z_{3} & a & b\\
Z_{4} & c & d\end{array}\right)\label{eq:Short standard form}\end{equation}
where each of the (Hadamard) matrices $Z_{i}$ is fully determined
by a single complex number $z_{i}$ of modulus one, $|z_{i}|=1$,\begin{equation}
\begin{array}{ccccccc}
Z_{1} & = & \left(\begin{array}{cc}
1 & 1\\
z_{1} & -z_{1}\end{array}\right) & \hspace{8ex} & Z_{2} & = & \left(\begin{array}{cc}
1 & 1\\
z_{2} & -z_{2}\end{array}\right)\\
\\Z_{3} & = & \left(\begin{array}{cc}
1 & z_{3}\\
1 & -z_{3}\end{array}\right) &  & Z_{4} & = & \left(\begin{array}{cc}
1 & z_{4}\\
1 & -z_{4}\end{array}\right)\end{array},\label{eq:Z1Z2Z3Z4}\end{equation}
and where $a,\, b,\, c$ and $d$ are Hadamard matrices of order 2.
Not all matrices of the general form (\ref{eq:Short standard form})
will be Hadamard, and the additional conditions on the matrix elements
will now be investigated.

\section{The unitarity constraints}

In order to develop an exhaustive parametrization of the $H_{2}$-reducible
Hada\-mard matrices on the standard form (\ref{eq:Short standard form}),
the unitarity constraints on $H$ and its submatrices are first explored.
In a second step, the additional constraints imposed by the unimodularity
of the elements of $H$ are investigated. 

Let $e$ be the unit matrix in two dimensions.
\begin{prop}
\label{pro:ABCD}Let $H$ be an $H_{2}$-reducible Hadamard matrix
on the standard form (\ref{eq:Short standard form}). Then\[
H=\left(\begin{array}{ccc}
F_{2} & Z_{1} & Z_{2}\\
\\Z_{3} & \,\,\,\frac{1}{2}Z_{3}AZ_{1}\,\,\, & \frac{1}{2}Z_{3}BZ_{2}\\
\\Z_{4} & \frac{1}{2}Z_{4}BZ_{1} & \frac{1}{2}Z_{4}AZ_{2}\end{array}\right)\]
where\begin{eqnarray*}
A & = & F_{2}(-\frac{1}{2}e+i\frac{\sqrt{3}}{2}\Lambda)\\
B & = & F_{2}(-\frac{1}{2}e-i\frac{\sqrt{3}}{2}\Lambda)\end{eqnarray*}
and where the $2\times2$ matrix $\Lambda$ is unitary, $\Lambda^{\dagger}\Lambda=\Lambda\Lambda^{\dagger}=e$,
and self-adjoint, $\Lambda^{\dagger}=\Lambda$.\end{prop}
\begin{proof}
In (\ref{eq:Short standard form}), let $a=\frac{1}{2}Z_{3}AZ_{1}$,
$b=\frac{1}{2}Z_{3}BZ_{2}$, $c=\frac{1}{2}Z_{4}CZ_{1}$ and $d=\frac{1}{2}Z_{4}DZ_{2}$.
In terms of $A$, $B$, $C$ and $D$, the full set of unitarity constraints
on $H$ take the form\begin{equation}
\left\{ \begin{array}{ccc}
A & + & B=-F_{2}\\
C & + & D=-F_{2}\\
A & + & C=-F_{2}\\
B & + & D=-F_{2}\end{array}\right.\label{eq:LinearUnit}\end{equation}
and\begin{equation}
\left\{ \begin{array}{ccc}
AA^{\dagger}+BB^{\dagger} & = & 4e\\
CC^{\dagger}+DD^{\dagger} & = & 4e\\
AC^{\dagger}+BD^{\dagger} & = & -2e\end{array}\right.\hspace{8ex}\left\{ \begin{array}{ccc}
A^{\dagger}A+C^{\dagger}C & = & 4e\\
B^{\dagger}B+D^{\dagger}D & = & 4e\\
A^{\dagger}B+C^{\dagger}D & = & -2e\end{array}\right.\label{eq:QuadUnit}\end{equation}
Note that these conditions are independent of $z_{1},\, z_{2},\, z_{3}$
and $z_{4}$. It follows from (\ref{eq:LinearUnit}) that $D=A$ and
$C=B$. The relations (\ref{eq:QuadUnit}) can therefore be reduced
to\begin{equation}
\left\{ \begin{array}{ccc}
(A+B)(A+B)^{\dagger} & = & 2e\\
(A+B)^{\dagger}(A+B) & = & 2e\\
(A-B)(A-B)^{\dagger} & = & 6e\\
(A-B)^{\dagger}(A-B) & = & 6e\end{array}\right.\label{eq:QuadUnit 2}\end{equation}
\\
In view of the constraint $A+B=-F_{2}$ (from (\ref{eq:LinearUnit})),
the first two of these relations are always satisfied. In terms of
$\Lambda\equiv-iF_{2}(A-B)/(2\sqrt{3})$, the last two relations imply
that $\Lambda$ is unitary, $\Lambda^{\dagger}\Lambda=\Lambda\Lambda^{\dagger}=e$.
Finally, by assumption, $a^{\dagger}a=b^{\dagger}b=2e$ so that $A^{\dagger}A=B^{\dagger}B=2e$.
As a result, $(A+B)^{\dagger}(A-B)+(A-B)^{\dagger}(A+B)=0$ or, in
terms of $\Lambda$, $\Lambda-\Lambda^{\dagger}=0$. Solving for $A$
and $B$ in terms of $F_{2}$ and $\Lambda$ completes the proof.\end{proof}
\begin{lem}
\label{lem:Lambda}If a $2\times2$ matrix $\Lambda$ is unitary and
self-adjoint, either $\Lambda=\pm e$ or\[
\Lambda=\left(\begin{array}{cc}
\Lambda_{11} & \Lambda_{12}\\
\bar{\Lambda}_{12} & -\Lambda_{11}\end{array}\right)=\left(\begin{array}{cc}
\cos\theta & e^{i\phi}\sin\theta\\
e^{-i\phi}\sin\theta & -\cos\theta\end{array}\right)\]
with $\theta\in[0,2\pi)$ and $\phi\in[0,2\pi)$.\end{lem}
\begin{proof}
Since $\Lambda$ is self-adjoint, its diagonal elements $\Lambda_{11}$
and $\Lambda_{22}$ are real, and $\Lambda_{21}=\bar{\Lambda}_{12}$.
Furthermore, since $\Lambda$ is unitary,\[
\Lambda^{\dagger}\Lambda=\left(\begin{array}{cc}
\Lambda_{11}^{2}+|\Lambda_{12}|^{2} & \Lambda_{12}(\Lambda_{11}+\Lambda_{22})\\
\bar{\Lambda}_{12}(\Lambda_{11}+\Lambda_{22}) & \Lambda_{22}^{2}+|\Lambda_{12}|^{2}\end{array}\right)=\left(\begin{array}{cc}
1 & 0\\
0 & 1\end{array}\right).\]
\\
The off-diagonal elements vanish if either $\Lambda_{22}=-\Lambda_{11}$
or $\Lambda_{12}=\Lambda_{21}=0$. In the first case $\Lambda$ is
traceless, with $\Lambda_{11}^{2}+|\Lambda_{12}|^{2}=1$, i.e. $\Lambda$
can be parametrized as\[
\Lambda=\left(\begin{array}{cc}
\Lambda_{11} & \Lambda_{12}\\
\bar{\Lambda}_{12} & -\Lambda_{11}\end{array}\right)=\left(\begin{array}{cc}
\cos\theta & e^{i\phi}\sin\theta\\
e^{-i\phi}\sin\theta & -\cos\theta\end{array}\right)\]
with $\theta\in[0,2\pi)$ and $\phi\in[0,2\pi)$, and $\det\Lambda=-1$.
In the second case $\Lambda$ is diagonal with $\Lambda_{11}^{2}=\Lambda_{22}^{2}=1$.
The possibility that $\Lambda_{11}=-\Lambda_{22}$ is already included
in the first case, leaving $\Lambda=\pm e$ as the only new possibilities,
and for which $\det\Lambda=1$.\end{proof}
\begin{rem*}
In more general terms, if a $2\times2$ unitary matrix $\Lambda$
is selfadjoint, either $\Lambda\subset SU(2)$, or $i\Lambda\subset SU(2)$.
In particular, the parametrization for $\Lambda$ given in Lemma \ref{lem:Lambda}
is directly related to the standard parametrization of $SU(2)$ matrices. \end{rem*}
\begin{cor}
\label{cor:AB}The matrices $A$ and $B$ of Proposition \ref{pro:ABCD}
either have the form (for $\Lambda=e$)\begin{equation}
A=\omega F_{2}\hspace{10ex}\mathrm{and}\hspace{10ex}B=\omega^{2}F_{2}\label{eq:AB case 1}\end{equation}
or (for $\Lambda=-e$)\begin{equation}
A=\omega^{2}F_{2}\hspace{10ex}\mathrm{and}\hspace{10ex}B=\omega F_{2}\label{eq:AB case 2}\end{equation}
with $\omega=-1/2+i\sqrt{3}/2=\exp(2\pi i/3),$ or otherwise (for
$\Lambda\ne\pm e$)\begin{equation}
A=\left(\begin{array}{cc}
A_{11} & A_{12}\\
\bar{A}_{12} & -\bar{A}_{11}\end{array}\right)\hspace{4ex}\mathrm{and}\hspace{4ex}B=\left(\begin{array}{cc}
B_{11} & B_{12}\\
\bar{B}_{12} & -\bar{B}_{11}\end{array}\right)\label{eq:AB case 3}\end{equation}
where\begin{eqnarray*}
A_{11} & = & -\frac{1}{2}+i\frac{\sqrt{3}}{2}(\,\,\cos\theta+e^{-i\phi}\,\sin\theta)\\
A_{12} & = & -\frac{1}{2}+i\frac{\sqrt{3}}{2}(-\cos\theta+e^{i\phi}\,\,\sin\theta)\end{eqnarray*}
and $B=-F_{2}-A$. 
\end{cor}
At this point, all unitarity constraints on the matrix $H$ and its
submatrices have been accounted for. Note that although the matrices
$A$ and $B$ satisfy the unitarity constraints, they will in general
not be Hadamard (the modulus of the matrix elements will not be equal
to one).

\section{\label{sec:The-unimodularity-constraints}The unimodularity constraints}

The additional condition that all elements of $H$ should be of unit
modulus can now be imposed. 
\begin{prop}
\label{pro:mod_cond}Let $H$ be an $H_{2}$-reducible Hadamard matrix
on the form (\ref{eq:Short standard form}), and let $A$ and $B$
be as in Proposition \ref{pro:ABCD} and Corollary \ref{cor:AB}.
For $A$ and $B$ according to (\ref{eq:AB case 1}) or (\ref{eq:AB case 2}),
the elements of $a$, $b$, $c$ and $d$ are of unit modulus if\begin{eqnarray}
(1-z_{1}^{2})(1-z_{3}^{2}) & = & 0\nonumber \\
(1-z_{2}^{2})(1-z_{4}^{2}) & = & 0\nonumber \\
(1-z_{1}^{2})(1-z_{4}^{2}) & = & 0\nonumber \\
(1-z_{2}^{2})(1-z_{3}^{2}) & = & 0.\label{eq:Hada_cond}\end{eqnarray}
For $A$ and $B$ according to (\ref{eq:AB case 3}), the elements
of $a$, $b$, $c$ and $d$ are of unit modulus if\begin{eqnarray}
-A_{11}^{2}+z_{1}^{2}A_{12}^{2}+z_{3}^{2}\bar{A}_{12}^{2}-z_{1}^{2}z_{3}^{2}\bar{A}_{11}^{2} & = & 0\nonumber \\
-B_{11}^{2}+z_{2}^{2}B_{12}^{2}+z_{3}^{2}\bar{B}_{12}^{2}-z_{2}^{2}z_{3}^{2}\bar{B}_{11}^{2} & = & 0\nonumber \\
-B_{11}^{2}+z_{1}^{2}B_{12}^{2}+z_{4}^{2}\bar{B}_{12}^{2}-z_{1}^{2}z_{4}^{2}\bar{B}_{11}^{2} & = & 0\nonumber \\
-A_{11}^{2}+z_{2}^{2}A_{12}^{2}+z_{4}^{2}\bar{A}_{12}^{2}-z_{2}^{2}z_{4}^{2}\bar{A}_{11}^{2} & = & 0.\label{eq:unimod cond gen}\end{eqnarray}
\end{prop}
\begin{proof}
The elements of $a=\frac{1}{2}Z_{3}AZ_{1}$ can all be expressed in
terms of \\  $a_{11}(z_{1},z_{3})=(A_{11}+z_{1}A_{12}+z_{3}A_{21}+z_{1}z_{3}A_{22})/2$,\[
a=\left(\begin{array}{cc}
a_{11}(z_{1},z_{3})\,\, & \,\, a_{11}(-z_{1},z_{3})\\
a_{11}(z_{1},-z_{3})\,\, & \,\, a_{11}(-z_{1},-z_{3})\end{array}\right).\]
For $A$ and $B$ according to (\ref{eq:AB case 1}) or (\ref{eq:AB case 2}),
the four conditions \[
|a_{11}(\pm z_{1},\pm z_{3})|^{2}=1\]
 all reduce to the first of the relations (\ref{eq:Hada_cond}), while
for $A$ and $B$ according to (\ref{eq:AB case 3}), the first of
the relations (\ref{eq:unimod cond gen}) is obtained. The remaining
relations follow in a similar manner by considering $b$, $c$ and
$d$. 
\end{proof}
With this results, all the conditions needed to characterize the set
of $H_{2}$-reducible Hadamard matrices have been given in an explicit
form. Before examining these conditions in detail, however, some additional
constraints will be imposed that come from the desire to obtain a
characterization in terms of inequivalent matrices.
\begin{prop}
\label{pro:z sign ambig}Given $A$ and $B$, the 16 possible sign
combinations for the $z_{i}$ parameters obtained when solving (\ref{eq:Hada_cond})
or (\ref{eq:unimod cond gen}) generate equivalent sets of Hadamard
matrices.\end{prop}
\begin{proof}
The conditions (\ref{eq:Hada_cond}) and (\ref{eq:unimod cond gen})
only determine the $z_{i}$ parameters up to a sign. However, a sign
change can be compensated by an interchange of rows and/or of columns,
and the resulting Hadamard matrix is therefore equivalent to the original
one. For instance, let $H'$ and $H''$ only differ in the sign of
$z_{3}$,\[
Z_{3}'=\left(\begin{array}{cc}
1 & z_{0}\\
1 & -z_{0}\end{array}\right)\,\,\,\,\,\mathrm{and}\,\,\,\,\, Z_{3}''=\left(\begin{array}{cc}
1 & -z_{0}\\
1 & z_{0}\end{array}\right)=PZ_{3}',\]
where $P=\left(\begin{array}{cc}
0 & 1\\
1 & 0\end{array}\right)$ is a row-permuting matrix. Then\[
H''=\left(\begin{array}{ccc}
F_{2} & Z_{1} & Z_{2}\\
Z_{3}'' & \,\,\,\frac{1}{2}Z_{3}''AZ_{1}\,\,\, & \frac{1}{2}Z_{3}''BZ_{2}\\
Z_{4} & \frac{1}{2}Z_{4}BZ_{1} & \frac{1}{2}Z_{4}AZ_{2}\end{array}\right)=\left(\begin{array}{ccc}
e & 0 & 0\\
0 & P & 0\\
0 & 0 & e\end{array}\right)H'\sim H'\]

\end{proof}
As a consequence of Proposition \ref{pro:z sign ambig}, in order
to map out the family of all non-equivalent $H_{2}$-reducible Hadamard
matrices, only one sign for the $z_{i}$ parameters needs to be considered. 

It can also be shown that without loosing inequivalent matrices the
range of the $\theta$ and $\phi$ parameters of Lemma \ref{lem:Lambda}
can be reduced to $[0,\pi)$, and the special cases corresponding
to $\Lambda=\pm e$ (i.e. to Eqns (\ref{eq:AB case 1}) and (\ref{eq:AB case 2}))
can be disregarded. 
\begin{prop}
\label{pro:ABCD_theta_phi}Any $H_{2}$-reducible Hadamard matrix
is equivalent to a matrix on the form specified in Proposition \ref{pro:ABCD},
with\[
\Lambda=\left(\begin{array}{cc}
\Lambda_{11} & \Lambda_{12}\\
\bar{\Lambda}_{12} & -\Lambda_{11}\end{array}\right)=\left(\begin{array}{cc}
\cos\theta & e^{i\phi}\sin\theta\\
e^{-i\phi}\sin\theta & -\cos\theta\end{array}\right)\]
for $\theta\in[0,\pi)$, $\phi\in[0,\pi)$. \end{prop}
\begin{proof}
From Proposition \ref{pro:ABCD} it follows that a change of sign
$\Lambda\to-\Lambda$ induces the interchange $A\leftrightarrow B$,
\[
H\to H'=\left(\begin{array}{ccc}
F_{2} & Z_{1}' & Z_{2}'\\
\\Z_{3}' & \,\,\,\frac{1}{2}Z_{3}'BZ_{1}'\,\,\, & \frac{1}{2}Z_{3}'AZ_{2}'\\
\\Z_{4}' & \frac{1}{2}Z_{4}'AZ_{1}' & \frac{1}{2}Z_{4}'BZ_{2}'\end{array}\right)\]
When the interchange $A\leftrightarrow B$ is carried out in (\ref{eq:unimod cond gen}),
the resulting equations for the $z$-parameters are changed. If, however,
the original equations had solutions $z_{1}$, $z_{2}$, $z_{3}$
and $z_{4}$, the new equations will have solutions $z_{1}'=z_{1}$,
$z_{2}'=z_{2}$, $z_{3}'=z_{4}$ and $z_{4}'=z_{3}$. Therefore,\[
H'=\left(\begin{array}{ccc}
F_{2} & Z_{1} & Z_{2}\\
\\Z_{4} & \,\,\,\frac{1}{2}Z_{4}BZ_{1}\,\,\, & \frac{1}{2}Z_{4}AZ_{2}\\
\\Z_{3} & \frac{1}{2}Z_{3}AZ_{1} & \frac{1}{2}Z_{3}BZ_{2}\end{array}\right)\sim H\]
where in the last step some rows have been permuted. Therefore, in
order to map out the family of all non-equivalent $H_{2}$-reducible
Hadamard matrices, only one sign for $\Lambda$ needs to be considered. 

For the $\Lambda\ne\pm e$ case, the transformations $(\theta,\phi)\to(\theta+\pi,\phi)$
and $(\theta,\phi)\to(\pi-\theta,\phi+\pi)$ both imply $\Lambda\to-\Lambda$.
As a result, the range for $\theta$ and $\phi$ can be reduced to
$[0,\pi)$ . 

For the $\Lambda=\pm e$ case, only $\Lambda=e$ needs to be considered
further, and it will first be shown that the resulting Hadamard family
is equivalent to either of the two Fourier families. Indeed, from
(\ref{eq:Hada_cond}), either $z_{3}^{2}=z_{4}^{2}=1$, with $z_{1}^{2}$
and $z_{2}^{2}$ unconstrained, or $z_{1}^{2}=z_{2}^{2}=1$, with
$z_{3}^{2}$ and $z_{4}^{2}$ unconstrained. In the first case, let
$z_{3}=z_{4}=$1, so that $Z_{3}=Z_{4}=F_{2}$. The resulting Hadamard
matrices (see Corollary \ref{cor:AB})\[
H=\left(\begin{array}{ccc}
F_{2} & Z_{1} & Z_{2}\\
F_{2} & \omega Z_{1} & \omega^{2}Z_{2}\\
F_{2} & \omega^{2}Z_{1} & \omega Z_{2}\end{array}\right)\sim F_{6}^{(2)}\]
build the Fourier family, $F_{6}^{(2)}$, with $z_{1}$ and $z_{2}$
as parameters. In the second case, let $z_{1}=z_{2}=$1, so that $Z_{1}=Z_{2}=F_{2}$,
and the resulting matrices build the Fourier transposed family ($F_{6}^{(2)})^{T}$,
with $z_{3}$ and $z_{4}$ as parameters,\[
H=\left(\begin{array}{ccc}
F_{2} & F_{2} & F_{2}\\
Z_{3} & \omega Z_{3} & \omega^{2}Z_{3}\\
Z_{4} & \omega^{2}Z_{4} & \omega Z_{4}\end{array}\right)\sim(F_{6}^{(2)})^{T}.\]
However, as will be seen in the next section, $F_{6}^{(2)}$ and ($F_{6}^{(2)})^{T}$
also appear as limit families in the $\Lambda\ne\pm e$ case, for
$\theta\to0$ and $\theta\to\pi/2$. For the purpose of classifying
all $H_{2}$-reducible Hadamard matrices, the $\Lambda=\pm e$ cases
can therefore be disregarded from now on. 
\end{proof}

\section{\label{sec:The-three-parameter-family}The three-parameter family}

Given the matrices $A$ and $B$ of Proposition \ref{pro:ABCD}, or
more precisely the parameters $\theta$ and $\phi$ of Proposition
\ref{pro:ABCD_theta_phi}, what remains is to determine in detail
the conditions on the parameters $z_{i}$ that follow from the unimodularity
constraints (\ref{eq:unimod cond gen}). It is useful to see these
constraints as M\"obius transformations\emph{ }\[
w=\mathcal{M}(z)=\frac{\alpha z-\beta}{\bar{\beta}z-\bar{\alpha}},\hspace{8ex}z=\mathcal{M}^{-1}(w)=\frac{\bar{\alpha}w-\beta}{\bar{\beta}w-\alpha}\]
that, as long as $|\alpha|^{2}-|\beta|^{2}\neq0$, map the unit circle
onto itself. Formally, from (\ref{eq:unimod cond gen}),\begin{eqnarray}
z_{3}^{2}=\mathcal{M}_{A}(z_{1}^{2}) & \hspace{8ex} & z_{3}^{2}=\mathcal{M}_{B}(z_{2}^{2})\nonumber \\
z_{4}^{2}=\mathcal{M}_{A}(z_{2}^{2}) & \hspace{8ex} & z_{4}^{2}=\mathcal{M}_{B}(z_{1}^{2})\label{eq:Moebius}\end{eqnarray}
with $\alpha_{A}=A_{12}^{2}$, $\beta_{A}=A_{11}^{2}$, and $\alpha_{B}=B_{12}^{2}$,
$\beta_{B}=B_{11}^{2}$. Recall that the inverse of a M\"obius transformation,
as well as a sequence of two M\"obius transformations, is also a
M\"obius transformation. 

Through straightforward calculation, the following relation between
$\mathcal{M}_{A}$ and $\mathcal{M}_{B}$ can easily be verified (using
the expressions for $A$ and $B$ in terms of the $\Lambda$ of Propositions
\ref{pro:ABCD} and \ref{pro:ABCD_theta_phi}).
\begin{prop}
\label{pro:MM=00003DMM}For the M\"obius transformations of (\ref{eq:Moebius}),
\[
\mathcal{M}_{B}^{-1}(\mathcal{M}_{A}(z^{2}))=\mathcal{M}_{A}^{-1}(\mathcal{M}_{B}(z^{2})).\]

\end{prop}
In view of Proposition \ref{pro:MM=00003DMM}, the relations (\ref{eq:Moebius})
are not independent, but only allow for expressing three of the parameters
$z_{i}$ in terms of the fourth. Let for instance $z_{1}=\exp(i\psi_{1})$
where, considering Proposition \ref{pro:ABCD_theta_phi}, $\psi_{1}\in[0,\pi)$.
Then\begin{eqnarray*}
z_{3}^{2} & = & \mathcal{M}_{A}(z_{1}^{2})\\
z_{4}^{2} & = & \mathcal{M}_{B}(z_{1}^{2})\\
z_{2}^{2} & = & \mathcal{M}_{B}^{-1}(\mathcal{M}_{A}(z_{1}^{2}))=\mathcal{M}_{A}^{-1}(\mathcal{M}_{B}(z_{1}^{2}))\end{eqnarray*}
and the resulting set of Hadamard matrices will depend on the three
parameters $\theta,\,\phi$ and $\psi_{1}$. The same set will be
generated starting from any other $z_{i}$, and constitutes the advertised
three-parameter family of complex Hadamard matrices of order 6. 

The M\"obius transformations (\ref{eq:Moebius}) become degenerate
if $|\alpha|^{2}-|\beta|^{2}\to0$: the transformation $w=\mathcal{M}(z)$
degenerates into a mapping of the unit circle in $z$ into a single
point $w=\alpha/\bar{\beta}$, and this mapping has no inverse, and
the inverse transform $z=\mathcal{M}^{-1}(w)$ degenerates into a
mapping of the unit circle in $w$ into a single point $z=\bar{\alpha}/\bar{\beta}$,
and again there is no inverse. For $\mathcal{M}_{A}$ and $\mathcal{M}_{A}^{-1}$
this occurs if $|A_{11}|=|A_{12}|=1$, i.e. if\[
\sin\theta(\sin\phi-\sqrt{3}\cos\theta\cos\phi)=0,\]
and for $\mathcal{M}_{B}$ and $\mathcal{M}_{B}^{-1}$ if $|B_{11}|=|B_{12}|=1$,
i.e. if\[
\sin\theta(\sin\phi+\sqrt{3}\cos\theta\cos\phi)=0.\]
Both transformations are degenerate when $\theta=0$, any $\phi$
(and also when $\theta\to\pi$, any $\phi$), and when $\theta=\pi/2$,
$\phi=0$ (and also when $\theta=\pi/2$, $\phi\to\pi$). 

In general, such a degeneracy does not prevent the construction of
the three-parameter family as outlined above (see Appendix 1). However,
at the points where both transformations are degenerate, the analysis
must take into account that these points can be reached not only along
the degeneracy curves but from an arbitrary direction in the $\theta-\phi$
plane. The resulting limit families may be obtained either through
an explicit limiting procedure, as exemplified in Appendix 2, or in
the following direct manner. 

If $\mathcal{M}_{A}$ and $\mathcal{M}_{B}$ are both degenerate,
there are two cases to be considered. First, if $\theta=0$ then $\Lambda=\left(\begin{array}{cc}
1 & 0\\
0 & -1\end{array}\right)$ for any $\phi$, so that\[
A=F_{2}\Omega\hspace{3em}\mathrm{and}\hspace{3em}B=F_{2}\Omega^{2}\]
Here, $\Omega=\left(\begin{array}{cc}
\omega & 0\\
0 & \omega^{2}\end{array}\right)$ with $1+\omega+\omega^{2}=0$ and $e+\Omega+\Omega^{2}=0$ (recall
that $\omega=\exp(2\pi i/3)$). The unimodularity conditions (\ref{eq:unimod cond gen})
take the form\begin{eqnarray}
(z_{1}^{2}-\omega^{4})(z_{3}^{2}-1) & = & 0\nonumber \\
(z_{2}^{2}-\omega^{2})(z_{3}^{2}-1) & = & 0\nonumber \\
(z_{1}^{2}-\omega^{2})(z_{4}^{2}-1) & = & 0\nonumber \\
(z_{2}^{2}-\omega^{4})(z_{4}^{2}-1) & = & 0\label{eq:Hada_cond 0-0}\end{eqnarray}
This set requires that $z_{3}^{2}=1$ and/or $z_{4}^{2}=1$. If $z_{3}^{2}=z_{4}^{2}=1$,
then there are no restrictions on $z_{1}$ or $z_{2}$. Since all
sign combinations result in equivalent Hadamard matrices, let $z_{3}=z_{4}=1$.
Then $Z_{3}=Z_{4}=F_{2}$, and the resulting Hadamard family is equivalent
to the Fourier family $F_{6}^{(2)}$, \[
H=\left(\begin{array}{ccc}
F_{2} & Z_{1} & Z_{2}\\
F_{2} & \Omega Z_{1} & \Omega^{2}Z_{2}\\
F_{2} & \Omega^{2}Z_{1} & \Omega Z_{2}\end{array}\right)\sim\left(\begin{array}{ccc}
F_{2} & Z_{1} & Z_{2}\\
F_{2} & \omega Z_{1} & \omega^{2}Z_{2}\\
F_{2} & \omega^{2}Z_{1} & \omega Z_{2}\end{array}\right)\sim F_{6}^{(2)}.\]
The system (\ref{eq:Hada_cond 0-0}) is also satisfied if $z_{3}^{2}=1$,
$z_{1}^{2}=\omega^{2}$ and $z_{2}^{2}=\omega^{4}$, with $z_{4}$
arbitrary. Let $z_{3}=1,$ $z_{1}=\omega$ and $z_{2}=\omega^{2}$.
In this case\[
H=\left(\begin{array}{cccccc}
1 & 1 & 1 & 1 & 1 & 1\\
1 & -1 & \omega & -\omega & \omega^{2} & -\omega^{2}\\
1 & 1 & \omega & \omega & \omega^{2} & \omega^{2}\\
1 & -1 & 1 & -1 & 1 & -1\\
1 & z_{4} & \omega^{2} & \omega^{2}z_{4} & \omega & \omega z_{4}\\
1 & -z_{4} & \omega^{2} & -\omega^{2}z_{4} & \omega & -\omega z_{4}\end{array}\right)\sim\left(\begin{array}{ccc}
F_{2} & F_{2} & F_{2}\\
F_{2} & \omega F_{2} & \omega^{2}F_{2}\\
Z_{4} & \omega^{2}Z_{4} & \omega Z_{4}\end{array}\right)\]
and this family is equivalent to a subfamily of $(F_{6}^{(2)})^{T}$.
Finally, the system (\ref{eq:Hada_cond 0-0}) is also satisfied if
$z_{4}^{2}=1$, $z_{1}^{2}=\omega^{4}$ and $z_{2}^{2}=\omega^{2}$,
with $z_{3}$ arbitrary. Like in the previous case, the resulting
$H$ can be shown to be equivalent to a subfamily of $(F_{6}^{(2)})^{T}$. 

The M\"obius transformations $\mathcal{M}_{A}$ and $\mathcal{M}_{B}$
are also degenerate when $\theta=\pi/2$, $\phi=0$, and in this case
$\Lambda=\left(\begin{array}{cc}
0 & 1\\
1 & 0\end{array}\right)$ and\[
A=\Omega F_{2}\hspace{3em}\mathrm{and}\hspace{3em}B=\Omega^{2}F_{2}.\]
The subsequent analysis is similar to the previous one, and results
in matrix families that either are equivalent to  $(F_{6}^{(2)})^{T}$,
or to one-parameter subfamilies of $F_{6}^{(2)}$. 

The finding of two-parameter subfamilies at the doubly degenerate
points might not have been expected, since two ($\theta$ and $\phi$)
of the three original parameters have been eliminated. It might be
recalled, however, that a similar phenomenon was observed in \citep{Karlsson_JMP},
where the two-parameter family $K_{6}^{(2)}$ at certain fixed parameter
values had the one-parameter $D_{6}^{(1)}$ family as limit family.
As was detailed in \citep{Karlsson_JMP}, the extra parameter enters
since the limit family depends on the direction from which the limit
point is reached, just as is observed here (see Appendix 2). 

It should be recalled that the appearance of the Fourier and Fourier
transposed families in the present context was made use of in the
proof of Proposition \ref{pro:ABCD_theta_phi}.

With these observations, the classification problem for $H_{2}$-reducible
Hada\-mard matrices is solved. The main results of the present paper
are collected in the following theorem.
\begin{thm}
\label{thm:Main result}Any $H_{2}$-reducible (complex) Hadamard
matrices (of order 6) is equivalent to a member of the three-parameter
family of dephased matrices\[
H=\left(\begin{array}{ccc}
F_{2} & Z_{1} & Z_{2}\\
\\Z_{3} & \,\,\,\frac{1}{2}Z_{3}AZ_{1}\,\,\, & \frac{1}{2}Z_{3}BZ_{2}\\
\\Z_{4} & \frac{1}{2}Z_{4}BZ_{1} & \frac{1}{2}Z_{4}AZ_{2}\end{array}\right)\]
Here $F_{2}=\left(\begin{array}{cc}
1 & 1\\
1 & -1\end{array}\right)$, $A$ is the matrix\[
A=\left(\begin{array}{cc}
A_{11} & A_{12}\\
\bar{A}_{12} & -\bar{A}_{11}\end{array}\right)\]
with elements\begin{eqnarray*}
A_{11} & = & -\frac{1}{2}+i\frac{\sqrt{3}}{2}(\,\,\cos\theta+e^{-i\phi}\,\sin\theta)\\
A_{12} & = & -\frac{1}{2}+i\frac{\sqrt{3}}{2}(-\cos\theta+e^{i\phi}\,\,\sin\theta)\end{eqnarray*}
for any $\theta\in[0,\pi)$ and $\phi\in[0,\pi)$, and $B=-F_{2}-A$.
In the matrices $Z_{i}=\left(\begin{array}{cc}
1 & 1\\
z_{i} & -z_{i}\end{array}\right)$, $i=1,2$, and $Z_{i}=\left(\begin{array}{cc}
1 & z_{i}\\
1 & -z_{i}\end{array}\right)$, $i=3,4$, the parameters $z_{i}$ are related through M\"obius
transformations\begin{eqnarray*}
z_{3}^{2}=\mathcal{M}_{A}(z_{1}^{2}) & \hspace{8ex} & z_{3}^{2}=\mathcal{M}_{B}(z_{2}^{2})\\
z_{4}^{2}=\mathcal{M}_{A}(z_{2}^{2}) & \hspace{8ex} & z_{4}^{2}=\mathcal{M}_{B}(z_{1}^{2})\end{eqnarray*}
where\[
w=\mathcal{M}(z)=\frac{\alpha z-\beta}{\bar{\beta}z-\bar{\alpha}}\]
with $\alpha_{A}=A_{12}^{2}$, $\beta_{A}=A_{11}^{2}$, and $\alpha_{B}=B_{12}^{2}$,
$\beta_{B}=B_{11}^{2}$. In general, one of the parameters $z_{i}$
can be chosen freely, say $z_{1}=\exp(i\psi_{1})$, $\psi_{1}\in[0,\pi)$,
in which case $z_{2}^{2}=\mathcal{M}_{A}^{-1}(\mathcal{M}_{B}(z_{1}^{2}))=\mathcal{M}_{B}^{-1}(\mathcal{M}_{A}(z_{1}^{2}))$,
$z_{3}^{2}=\mathcal{M}_{A}(z_{1}^{2})$ and $z_{4}^{2}=\mathcal{M}_{B}(z_{1}^{2})$.
Any sign combinations for $z_{1}$, $z_{2}$, $z_{3}$ and $z_{4}$
lead to three-parameter families that are equivalent to each other. 
\end{thm}

\section{Special cases}

As pointed out above, all so far (analytically) known one- and two-parameter
families of complex Hadamard matrices of order 6 are subfamilies of
the three-parameter family constructed in the previous sections. In
general, however, the parameters used to classify these subfamilies
differ from the parameters introduced here, and the detailed connection
is not always transparent. For instance, the two-parameter family
$K_{6}^{(2)}$ of \citep{Karlsson_JMP} exploits simplifications entailed
by the assumption that $z_{2}=z_{1}$ and $z_{4}=z_{3}$. Such an
assumption is less natural from the point of view of the parametrization
developed in the present paper, and amounts to introducing a dependence
between $z_{1}$ and the parameters $\theta$ and $\phi$. In this
respect, the family $D_{6}^{(1)}$ is an exception, as will be shown
next.

Particularly simple subfamilies of the three-parameter family can
be expected if $\theta$ and $\phi$ kept constant. Consider for example
the point $\theta=\arccos(1/\sqrt{3})$, $\phi=\pi/4$, for which\[
\left\{ \begin{array}{ccc}
A_{11} & = & i\\
A_{12} & = & -1\end{array}\right.\hspace{3em}\mathrm{and}\hspace{3em}\left\{ \begin{array}{ccc}
B_{11} & = & -1-i\\
B_{12} & = & 0\end{array}\right.\]
Since $|A_{11}|=|A_{12}|=1$, $\mathcal{M}_{A}$ is degenerate and
$\mathcal{M}_{A}(z^{2})=\mathcal{M}_{A}^{-1}(z^{2})=-1$. Furthermore,
$\mathcal{M}_{B}(z^{2})=\mathcal{M}_{B}^{-1}(z^{2})=1/z^{2}=\bar{z}^{2}$
so that, taking $z_{1}=z$ as independent parameter, $z_{2}^{2}=-1$,
$z_{3}^{2}=-1$ and $z_{4}^{2}=\bar{z}^{2}$. Let $z_{2}=z_{3}=i$
and $z_{4}=\bar{z}$. The resulting one-parameter Hadamard matrix\[
H=\left(\begin{array}{cccccc}
1 & \,\,\,1 & \,\,\,1 & \,\,\,1 & \,\,\,1 & \,\,\,1\\
1 & -1 & \,\,\, z & -z & \,\,\, i & -i\\
1 & \,\,\, i & -z & \,\,\, z & -1 & -i\\
1 & -i & \,\,\, i & \,\,\, i & -i & -1\\
1 & \,\,\,\bar{z} & -i & -1 & -\bar{z} & \,\,\, i\\
1 & -\bar{z} & -1 & -i & \,\,\,\bar{z} & \,\,\, i\end{array}\right)\]
is equivalent to the generic member of $D_{6}^{(1)}$. 

Another example of a simple subfamily can be obtained as follows.
For points on the $\mathcal{M}_{A}$ degeneracy curve, $\sin\phi=\sqrt{3}\cos\theta\cos\phi$
(see Section \ref{sec:The-three-parameter-family}). Along this curve
$A_{12}=\exp(2i\phi)A_{11}$ and $\mathcal{M}_{A}^{-1}(z^{2})=\exp(-4i\phi)$,
all $z^{2}$. Let $z_{1}=z=\exp(i\psi)$. If $\phi$ is chosen equal
to $-\psi/2$ then $z_{2}^{2}=\mathcal{M}_{A}^{-1}(z^{2})=z^{2}$,
i.e. $z_{1}=z_{2}=z$. Furthermore \[
z_{3}=z_{4}=\frac{1-i\epsilon\sqrt{1+z+\bar{z}}}{1+i\epsilon\sqrt{1+z+\bar{z}}}\]
for $\psi\in[0,2\pi/3]$, and the resulting Hadamard matrix is equivalent
to \[
\left(\begin{array}{cccccc}
1 & \,\,\,1 & \,\,\,1 & \,\,\,1 & \,\,\,1 & \,\,\,1\\
1 & -1 & \,\,\, z & -z & \,\,\, z & -z\\
1 & \,\,\, z_{3} & \,\,\, z_{3}z & \,\,\, z & -\sqrt{z_{3}z} & -\sqrt{z_{3}z}\\
1 & -z_{3} & \,\,\, z_{3} & -1 & -\sqrt{z_{3}z} & \,\,\,\sqrt{z_{3}z}\\
1 & \,\,\, z_{3} & -\sqrt{z_{3}z} & \,\,-\sqrt{z_{3}z} & \,\,\, z_{3}z & \,\,\, z\\
1 & -z_{3} & -\sqrt{z_{3}z} & \,\,\,\,\,\sqrt{z_{3}z} & \,\,\, z_{3} & -1\end{array}\right)\]
This one-parameter Hadamard family can be identified as a subfamily
of $K_{6}^{(2)}$.

\section{Summary and outlook}

With the results of the present paper, the characterization problem
for complex Hadamard matrices of order six has been given a partial
solution, in that the subset of $H_{2}$-reducible Hadamard matrices
has been fully described in terms of a single, three-parameter family.
There is strong numerical evidence, based on some $10^{5}$ non-reducible
Hadamard matrices (see also \citep{Skinner}) that a full characterization
requires an additional parameter, but it remains an open question
whether or not closed form expressions for such a four-parameter family
can be found.

The parameters chosen here for the three-parameter family are not
unique, but appear as a natural choice. Minor variations, like choosing
the (SU(2)) parameters $\theta$ and $\phi$ differently, offer no
obvious advantage. 

As an application of the results presented here, Hadamard matrices
in 12 dimensions can be constructed. Such an extension was outlined
in \citep{Karlsson_JMP} based on the at the time known two-parameter
families in six dimensions. A corresponding extension based on the
three-parameter family of the present paper results in an eleven-parameter
family, the largest family constructed so far in 12 dimensions.

\section*{Appendix 1: Degenerate transformations}

If one of the M\"obius transformations (\ref{eq:Moebius}) becomes
degenerate, the three-parameter family may still be constructed as
outlined in Section \ref{sec:The-three-parameter-family}, but the
result may depend on how the degeneracy limit is approached. In order
to illustrate this point, let $\mathcal{M}_{A}$ but not $\mathcal{M}_{B}$
be degenerate. In such a case, $\mathcal{M}_{A}(z^{2})=w_{0}^{2}$
and $\mathcal{M}_{A}^{-1}(z^{2})=z_{0}^{2}$ for any $z$, where $w_{0}^{2}=\alpha_{A}/\bar{\beta}_{A}$
and $z_{0}^{2}=\bar{\alpha}_{A}/\bar{\beta}_{A}$ are uniquely specified
by $\theta$ or $\phi$ along the degeneracy curve. Furthermore, $\mathcal{M}_{B}(z_{0}^{2})=w_{0}^{2}$
as a consequence of Proposition \ref{pro:MM=00003DMM}. Using $z_{1}$
or $z_{4}$ as independent parameter, the remaining parameters are
obtained through $z_{4}^{2}=\mathcal{M}_{B}(z_{1}^{2})$ with $z_{3}^{2}=w_{0}^{2}$
and $z_{2}^{2}=z_{0}^{2}$. On the other hand, taking $z_{2}$ or
$z_{3}$ as independent parameter leads to $z_{3}^{2}=\mathcal{M}_{B}(z_{2}^{2})$
with $z_{1}^{2}=z_{0}^{2}$ and $z_{4}^{2}=w_{0}^{2}$. The resulting
two limit matrices, \[
\left(\begin{array}{ccc}
F_{2} & Z_{1} & Z_{z_{0}}\\
\\Z_{w_{0}} & \frac{1}{2}Z_{w_{0}}AZ_{1} & \frac{1}{2}Z_{w_{0}}BZ_{z_{0}}\\
\\Z_{4} & \frac{1}{2}Z_{4}BZ_{1} & \frac{1}{2}Z_{4}AZ_{z_{0}}\end{array}\right)\mathrm{and}\left(\begin{array}{ccc}
F_{2} & Z_{z_{0}} & Z_{2}\\
\\Z_{3} & \frac{1}{2}Z_{3}AZ_{z_{0}} & \frac{1}{2}Z_{3}BZ_{2}\\
\\Z_{w_{0}} & \frac{1}{2}Z_{w_{0}}BZ_{z_{0}} & \frac{1}{2}Z_{w_{0}}AZ_{2}\end{array}\right)\]
in obvious notation, are seemingly different, but an interchange of
rows and of columns shows that they generate families of equivalent
Hadamard matrices. There is therefore no need to amend the general
construction in Section \ref{sec:The-three-parameter-family} with
additional rules when one of the M\"obius transformations becomes
degenerate.

\section*{Appendix 2: The limit families at $\theta=0$}

In order to see how the general, three-parameter family behaves when
the doubly degenerate point at $\theta=0$ is approached, let $\theta$
be infinitesimal in the expressions for $A$ and $B$ in Theorem \ref{thm:Main result},
\[
\begin{array}{ccccccc}
A_{11} & \approx & \omega+i\frac{\sqrt{3}}{2}e^{-i\phi}\theta & \,\,\,\,\,\,\,\,\,\,\,\,\,\,\, & B_{11} & \approx & \omega^{2}-i\frac{\sqrt{3}}{2}e^{-i\phi}\theta\\
A_{12} & \approx & \omega^{2}+i\frac{\sqrt{3}}{2}e^{i\phi}\theta &  & B_{12} & \approx & \omega-i\frac{\sqrt{3}}{2}e^{i\phi}\theta\end{array}\]
 where $\omega=\exp(2\pi i/3)$. The coefficients of the M\"obius
transformations (\ref{eq:Moebius}) are\[
\begin{array}{ccccccc}
\alpha_{A} & \approx & \omega+i\sqrt{3}\omega^{2}e^{i\phi}\theta & \,\,\,\,\,\,\,\,\,\,\,\,\,\,\, & \alpha_{B} & \approx & \omega^{2}-i\sqrt{3}\omega e^{i\phi}\theta\\
\beta_{A} & \approx & \omega^{2}+i\sqrt{3}\omega e^{-i\phi}\theta &  & \beta_{B} & \approx & \omega-i\sqrt{3}\omega^{2}e^{-i\phi}\theta\end{array}.\]
Choosing $z_{1}$ as the independent parameter results in the relations,
when $\theta\to0$,\begin{eqnarray*}
z_{3}^{2} & = & \mathcal{M}_{A}(z_{1}^{2})\to\left\{ \begin{array}{rcccc}
1 &  & z_{1}^{2} & \ne & \omega^{4}\\
-1 &  & z_{1}^{2} & = & \omega^{4}\end{array}\right.\\
z_{4}^{2} & = & \mathcal{M}_{B}(z_{1}^{2})\to\left\{ \begin{array}{rcccc}
1 &  & z_{1}^{2} & \ne & \omega^{2}\\
-1 &  & z_{1}^{2} & = & \omega^{2}\end{array}\right.\\
z_{2}^{2} & = & \mathcal{M}_{A}^{-1}(\mathcal{M}_{B}(z_{1}^{2}))\to-e^{2i\phi}\frac{(1+e^{-2i\phi})z_{1}^{2}+e^{-2i\phi}}{e^{2i\phi}z_{1}^{2}+1+e^{2i\phi}}\end{eqnarray*}
 Given $z_{1}$ (not equal to $\omega$ or $\omega^{2}$), $\phi$
maps out a unit circle in $z_{2}$, i.e. the Fourier family $F_{6}^{(2)}$
with $z_{1}$ and $z_{2}$ as independent parameters is obtained. 

On the other hand, choosing $z_{3}$ as independent parameter results
in \begin{eqnarray*}
z_{1}^{2} & = & \mathcal{M}_{A}^{-1}(z_{3}^{2})\to\left\{ \begin{array}{ccccc}
\omega &  & z_{3}^{2} & \ne & 1\\
\omega^{2}e^{-2i\phi} &  & z_{3}^{2} & = & 1\end{array}\right.\\
z_{2}^{2} & = & \mathcal{M}_{B}^{-1}(z_{3}^{2})\to\left\{ \begin{array}{ccccc}
\omega^{2} & \hspace{1ex} & z_{3}^{2} & \ne & 1\\
\omega e^{-2i\phi} &  & z_{3}^{2} & = & 1\end{array}\right.\\
z_{4}^{2} & = & \mathcal{M}_{A}(\mathcal{M}_{B}^{-1}(z_{3}^{2}))\to1\end{eqnarray*}
Therefore, any $z_{3}\ne1$ leaves the other three parameters fixed,
and, as detailed in section 6.3, the resulting Hadamard family is
equivalent to a subfamily of ($F_{6}^{(2)})^{T}$.

\end{document}